\documentclass[slac_one]{revtex4}
\usepackage{epsfig,amsmath,feynarts}
\usepackage{graphicx}
\usepackage{fancyhdr}
\newcommand{\sweff}{\sin^2\theta_{\mathrm{eff}}}

\newcommand{\gev}{\,\, \mathrm{GeV}}
\newcommand{\mev}{\,\, \mathrm{MeV}}
\def\order#1{${\cal O}(#1)$}

\begin{document}

\title{{\small{2005 International Linear Collider Workshop - Stanford,
U.S.A.}}\\ 
\vspace{12pt}
Two-loop contributions to electroweak precision observables 
in the MSSM} 

%

\author{J.~Haestier, D.~St\"ockinger, G.~Weiglein}
\affiliation{IPPP, 
University of Durham, Durham DH1 3LE, U.K.}
\author{S.~Heinemeyer}
\affiliation{CERN, TH Division, Dept. of Physics, 1211 Geneva 23,
  Switzerland}

\begin{abstract}

The evaluation of the two-loop MSSM-contributions to the electroweak
precision observables $M_W$ and $\sweff$ at 
\order{\alpha_t^2}, \order{\alpha_t  \alpha_b}, \order{\alpha_b^2} is
presented. These contributions enter via $\Delta\rho$, and it is
explained in detail how one can retain the true, 
non-vanishing value of the MSSM Higgs boson mass $M_h$ in spite
of using the gauge-less limit in the calculation. The numerical
results can be sizeable, in particular for 
strong squark mixing. By  comparing the results in the on-shell and
$\overline{\rm DR}$ renormalization schemes, the remaining theoretical
uncertainty is found to be small.

\end{abstract}

\maketitle

\thispagestyle{fancy}

\section{INTRODUCTION}

Electroweak precision observables (EWPO) like $M_W$ or the effective
leptonic weak mixing angle $\sweff$ are intimately related to the
quantum structure of the electroweak interactions. The experimental
determination of these quantities, obtained at LEP and Tevatron, has a
precision of better than one per-mille: $\delta M_W=34 \mev$ (0.04\%) and
$\delta\sweff=16\times10^{-5}$ (0.07\%) \cite{LEPEWWG}. In the future,
these accuracies will improve, and at the GigaZ option of a linear
$e^+e^-$ collider a precision of $\delta M_W=7
\mev$~\cite{mwgigaz,blueband} and
$\delta\sweff=1.3\times10^{-5}$~\cite{swgigaz,blueband} can be achieved.

These precise measurements constitute tests of the quantum level
of the Standard Model (SM) that are sensitive even to two-loop
effects. The corresponding theoretical evaluation of the SM
predictions up to the two-loop level is quite advanced, see in
particular Ref.\ \cite{Ayres} for the most recent developments.
On the other hand, the measurements of the EWPO can be used to
discriminate between different models of electroweak interactions and
to derive constraints on unknown parameters. The comparison of the SM
and its minimal supersymmetric extension, the MSSM, is particularly
interesting since the MSSM agrees with all precision data at least as
well as the SM, in some cases even better. 

\begin{figure}[tb]
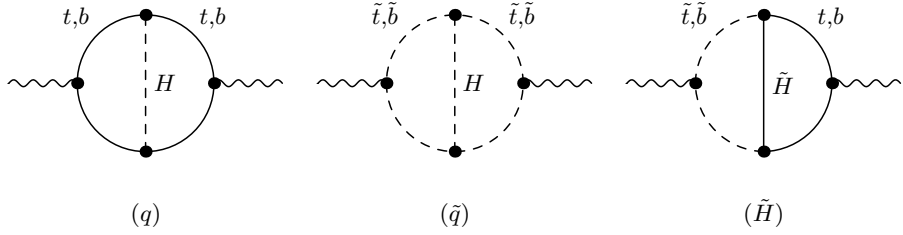

\centerline{
\unitlength=1.cm%
\begin{feynartspicture}(4,4)(1,1)
\FADiagram{$(q)$}
\FAVert(5,10){0}
\FAVert(15,10){0}
\FAProp(0,10)(5,10)(0.,){/Sine}{0}
\FAProp(15,10)(20,10)(0.,){/Sine}{0}
{
\FAProp(5,10)(15,10)(1.,){/Straight}{0}
\FAProp(15,10)(5,10)(1.,){/Straight}{0}
\FALabel(4,14)[b]{\ ${t}$,${b}$}
\FALabel(14,14)[b]{\ ${t}$,${b}$}}
\FAVert(10,15){0}
\FAVert(10,5){0}
\FAProp(10,5)(10,15)(0.,){/ScalarDash}{0}
\FALabel(11,10)[r]{$H$}
\end{feynartspicture}
\unitlength=1.cm%
\begin{feynartspicture}(4,4)(1,1)
\FADiagram{$(\tilde{q})$}
\FAVert(5,10){0}
\FAVert(15,10){0}
\FAProp(0,10)(5,10)(0.,){/Sine}{0}
\FAProp(15,10)(20,10)(0.,){/Sine}{0}
{
\FAProp(5,10)(15,10)(1.,){/ScalarDash}{0}
\FAProp(15,10)(5,10)(1.,){/ScalarDash}{0}
\FALabel(4,14)[b]{\ $\tilde{t}$,$\tilde{b}$}
\FALabel(14,14)[b]{\ $\tilde{t}$,$\tilde{b}$}}
\FAVert(10,15){0}
\FAVert(10,5){0}
\FAProp(10,5)(10,15)(0.,){/ScalarDash}{0}
\FALabel(11,10)[r]{$H$}
\end{feynartspicture}
\unitlength=1.cm%
\begin{feynartspicture}(4,4)(1,1)
\FADiagram{$(\tilde{H})$}
\FAVert(5,10){0}
\FAVert(15,10){0}
\FAProp(0,10)(5,10)(0.,){/Sine}{0}
\FAProp(15,10)(20,10)(0.,){/Sine}{0}
{
\FAProp(5,10)(10,15)(-.4,){/ScalarDash}{0}
\FAProp(10,5)(5,10)(-.4,){/ScalarDash}{0}
\FAProp(10,15)(15,10)(-.4,){/Straight}{0}
\FAProp(10,5)(15,10)(.4,){/Straight}{0}
\FALabel(4,14)[b]{\ $\tilde{t}$,$\tilde{b}$}
\FALabel(14,14)[b]{\ ${t}$,${b}$}}
\FAVert(10,15){0}
\FAVert(10,5){0}
\FAProp(10,5)(10,15)(0.,){/Straight}{0}
\FALabel(11,10)[r]{$\tilde{H}$}
\end{feynartspicture}
}
\caption{Sample diagrams for the three classes of contributions to
  $\Delta\rho$ considered here: $(q)$ $t/b$-quark loop with Higgs
  exchange, $(\tilde{q})$ $\tilde{t}/\tilde{b}$-squark loop with Higgs
  exchange, $(\tilde{H})$ mixed 
  quark/squark loop with Higgsino exchange.}
\label{fig:samplediagrams}
\end{figure}

It is therefore highly desirable to know the MSSM predictions for the
EWPO with a precision that matches the one of the SM and the
experiments. The comparison of the SM and the MSSM predictions
with the data could then lead to precise constraints e.g.\ on masses
of supersymmetric particles. As a step towards this goal we have
evaluated \cite{paper} the two-loop MSSM-corrections to the EWPO that
enter via $\Delta\rho$ at \order{\alpha_t^2}, \order{\alpha_t
  \alpha_b}, \order{\alpha_b^2}, where
\begin{equation}
\Delta\rho = \frac{\Sigma_Z(0)}{M_Z^2} - \frac{\Sigma_W(0)}{M_W^2} 
\label{delrho}
\end{equation}
in terms of the $Z$ and $W$ self energies $\Sigma_{Z,W}$. These are
leading two-loop contributions involving the top and bottom Yukawa
couplings and come from three classes of diagrams as shown in Fig.\
\ref{fig:samplediagrams}. These contributions to $\Delta\rho$ induce
universal two-loop corrections to the EWPO as follows (with $1-s_W^2 =
c_W^2 = M_W^2/M_Z^2$):
\begin{equation}
\delta M_W \approx \frac{M_W}{2}\frac{c_W^2}{c_W^2 - s_W^2} \Delta\rho, \quad
\delta\sweff \approx - \frac{c_W^2 s_W^2}{c_W^2 - s_W^2} \Delta\rho .
\label{precobs1}
\end{equation}
The previously known two-loop contributions to EWPO in the MSSM
comprise only QCD and SUSY-QCD corrections
\cite{dr2lA} and the \order{\alpha_t^2}, 
\order{\alpha_t \alpha_b}, \order{\alpha_b^2}
corrections for the class $(q)$ in Fig.\ \ref{fig:samplediagrams}
\cite{drMSSMgf2A}. Obviously, the diagrams of class $(q)$ contain no
supersymmetric particles. The contributions from classes 
$(\tilde{q})$, $(\tilde{H})$
considered here in addition can be expected to be important for
squark/Higgsino masses in the electroweak range and to have a more
pronounced dependence on the MSSM parameters. A similar situation was
found for the case of $(g-2)_\mu$, where the two-loop contributions
from squark-Higgs diagrams are more important than the ones from
quark-Higgs diagrams \cite{g-2}.

In the following we will first discuss in detail the renormalization
and the restrictions imposed by approximating the EWPO by $\Delta\rho$
as in eq.\ (\ref{precobs1}). It turns out that a strict calculation 
of $\Delta\rho$ at \order{\alpha_t^2}, 
\order{\alpha_t \alpha_b}, \order{\alpha_b^2} would imply a vanishing
MSSM Higgs boson mass, $M_h=0$, which would lead to a bad
approximation for the EWPO. We will show that it is possible to
improve the approximation by taking into account the true value of
$M_h$ essentially everywhere. Finally we will discuss the numerical
results and give an account of the remaining theoretical uncertainty
based on the dependence on the renormalization scheme.

\section{THE ROLE OF THE GAUGE-LESS LIMIT AND THE LIGHT HIGGS BOSON
  MASS \boldmath{$M_h$}}

\subsection{Gauge-less limit and \boldmath{$M_h=0$}}

$\Delta\rho$ as defined in eq.\ (\ref{delrho}) constitutes a part of
the loop corrections to the EWPO but is no observable. By itself, 
$\Delta\rho$ is only UV-finite and gauge independent if corrections
are considered that enter the EWPO {\em only} through vector boson
self energies in the form (\ref{precobs1}) (as opposed to e.g.\ vertex
or box diagrams). This is the case for the
\order{\alpha_t^2}, \order{\alpha_t \alpha_b}, \order{\alpha_b^2}
corrections.

Taking into account strictly only terms of \order{\alpha_t^2}, 
\order{\alpha_t \alpha_b}, \order{\alpha_b^2} corresponds in
particular to neglecting the gauge coupling $\alpha$ and thus to
taking the {\em gauge-less limit} $\alpha\to0$. In this limit,
$M_{Z,W}\to0$ while the ratio $c_W=M_W/M_Z$ is fixed.

In the SM the gauge-less limit is a reasonable approximation since
$\alpha\ll\alpha_t$. In the MSSM, however, the gauge-less limit has
side-effects in the Higgs sector since supersymmetry relates the Higgs
self couplings to gauge couplings. At tree-level, the gauge-less limit
implies the relations
\begin{subequations}
\label{glessrelations}
\begin{eqnarray}
\label{Mh0zero}
&&M_h=0,\\
&&M_{H^\pm}^2 = M_H^2 = M_A^2, \
\label{Mh0nonzero}
\sin\alpha = -\cos\beta ,\ \cos\alpha = \sin\beta
\end{eqnarray}
\end{subequations}
for the light and heavy CP-even, the CP-odd and the charged Higgs
boson masses 
$M_{h,H},M_A,M_{H^\pm}$ and the angles $\alpha,\beta$ where
$\tan\beta=v_2/v_1$, the ratio of the two vacuum expectation values.

The particularly troublesome of these relations is the masslessness of
the light Higgs boson, $M_h=0$. In the SM, where the Higgs boson
mass is a free parameter even in the gauge-less limit, one knows that
the result for $\Delta\rho$ at 
\order{\alpha_t^2}, \order{\alpha_t \alpha_b}, \order{\alpha_b^2}
is proportional to a factor \cite{drSMgf2mh0,drSMgf2mt4}
\begin{equation}
19 - 2 \pi^2 \mbox{ for }M_H=0,\quad
19 - 2 \pi^2 +f(M_H) \mbox{ for }M_H\ne0.
\end{equation}
For typical values of $M_H =$\order{100\gev}, this factor is
about an order of magnitude larger than for $M_H=0$, and the
result for $M_H=0$ leads to a bad approximation for the EWPO.

Due to the similar structure of the SM and MSSM diagrams it can be
expected that taking the gauge-less limit relation $M_h=0$ literally
would also lead to a bad approximation for the EWPO and should
therefore be avoided. Indeed, in Ref.\ \cite{drMSSMgf2A} it was
observed that the class $(q)$ contributions to $\Delta\rho$ in the
MSSM are already UV-finite if all relations in eq.\
(\ref{Mh0nonzero}) {\em but not} $M_h=0$ are employed. In the
following, we give an explanation of this result and show how it
extends to the contributions of classes $(\tilde{q})$, $(\tilde{H})$.

\subsection{Comparison of the MSSM and a general two-Higgs doublet
  model}

In order to study these questions it is very useful to regard the MSSM
as a special case of a more general two-Higgs-doublet model (2HDM)
without supersymmetry relations for the couplings. $\Delta\rho$ can be
calculated at 
\order{\alpha_t^2}, \order{\alpha_t \alpha_b}, \order{\alpha_b^2}
in the gauge-less limit in both models, but in a general 2HDM, the
gauge-less limit does not enforce any of the relations in eq.\
(\ref{glessrelations}). 

Comparing first the contributions from class $(q)$ in the MSSM and the
2HDM, we find that the corresponding two-loop diagrams and the
counterterm contributions from the top/bottom sector are identical. 
The only difference concerns the Higgs sector counterterm
contributions. In a general 2HDM, they can be derived from the
one-loop expression ($F_0$ is a symmetric function satisfying
$F_0(x,x) = 0$, $\partial_x F_0(x,y)|_{y=x}=0$)
\begin{equation}
\label{THDMHiggs}
\Delta\rho^{\rm H}_{\rm 2HDM} \propto
\Big[
F_0(M_{H^\pm}^2,M_{A^0}^2)
+s^2_{{\beta-\alpha}}\left(
F_0(M_{H^\pm}^2,M_{H}^2)-F_0(M_{A^0}^2,M_{H}^2)\right)
+ c^2_{\beta-\alpha}\left(
F_0(M_{H^\pm}^2,M_{h}^2)-F_0(M_{A^0}^2,M_{h}^2)\right)\Big]
\end{equation}
by performing the renormalization transformation $M_h\to M_h+\delta
M_h$ etc. In the MSSM, eq.\ (\ref{THDMHiggs}) and the Higgs sector
counterterms vanish because of the gauge-less limit relations
(\ref{glessrelations}). Thus the class $(q)$ contributions can be
decomposed as $\Delta\rho^{(q)}_{\rm 2-Loop}+
\Delta\rho^{(q)}_{\rm  t/b-cts}+\Delta\rho^{(q)}_{\rm H-cts}$ in the
2HDM and as $\Delta\rho^{(q)}_{\rm 2-Loop}+
\Delta\rho^{(q)}_{\rm  t/b-cts}$ in the MSSM. The 2HDM result is
UV-finite for all values of the Higgs sector parameters. 

From this
comparison we find that the MSSM result is not only UV-finite if
(\ref{glessrelations}) is used but more generally if
$\Delta\rho^{(q)}_{\rm H-cts}$ as derived from (\ref{THDMHiggs}) is
finite. From eq.\ (\ref{THDMHiggs}) we can explicitly read off that 
$\Delta\rho^{(q)}_{\rm H-cts}=0$ already if only the relations
(\ref{Mh0nonzero}) are used. This explains the observation made in
Ref.\ \cite{drMSSMgf2A}.
The MSSM result $\Delta\rho^{(q)}_{\rm 2-Loop}+
\Delta\rho^{(q)}_{\rm  t/b-cts}$ with the relations (\ref{Mh0nonzero}) 
corresponds to a certain special case of the general 2HDM
calculation. Therefore it is finite even for the true, non-vanishing
value of $M_h$.

For the class $(\tilde{q},\tilde{H})$ contributions we obtain similar
decompositions 
\begin{eqnarray}
\label{THDMRes}
\Delta\rho^{(\tilde{q},\tilde{H})}_{\rm 2HDM}&=&
\Delta\rho^{(\tilde{q},\tilde{H})}_{\rm 2-Loop}+
\Delta\rho^{(\tilde{q},\tilde{H})}_{\rm  t/b-cts}+
\Delta\rho^{(\tilde{q},\tilde{H},\rm OS4)}_{\rm
  \tilde{t}/\tilde{b}-cts}+
\Delta\rho^{(\tilde{q},\tilde{H})}_{\rm H-cts},\\
\Delta\rho^{(\tilde{q},\tilde{H})}_{\rm MSSM}&=&
\Delta\rho^{(\tilde{q},\tilde{H})}_{\rm 2-Loop}+
\Delta\rho^{(\tilde{q},\tilde{H})}_{\rm  t/b-cts}+
\Delta\rho^{(\tilde{q},\tilde{H},\rm OS3)}_{\rm
  \tilde{t}/\tilde{b}-cts}.
\label{MSSMRes}
\end{eqnarray}
In this case, even if the relations (\ref{Mh0nonzero}) are used such that
$\Delta\rho^{(\tilde{q},\tilde{H})}_{\rm H-cts}$ vanishes,  there is
still a difference between the 2HDM and the MSSM result because
the $\tilde{t}/\tilde{b}$ sector counterterms differ as indicated by
the superscripts OS4 and OS3.

In the MSSM, supersymmetry in conjunction with SU(2) gauge invariance
correlates the four sfermion masses $m_{\tilde{t}_{1,2}}$,
  $m_{\tilde{b}_{1,2}}$. Therefore  only three of them can be
renormalized independently. We choose to renormalize $m_{\tilde{t}_{1,2}}$,
and  $m_{\tilde{b}_{2}}$ independently by on-shell conditions (OS3). Then the fourth renormalization constant $\delta m_{\tilde{b}_1}$
is determined as a function of the other three in
the MSSM, while in the 2HDM
all four sfermion masses can be defined independently by 
on-shell conditions (OS4).
In the simple case of vanishing mixing between left- and right-handed
sfermions,  $m_{\tilde{b}_1}=m_{\tilde{b}_L}$ 
and  we obtain the following different results for $\delta
m_{\tilde{b}_L}^2$:
\begin{equation}
\delta m_{\tilde{b}_L}^2|_{\rm OS3} = \delta m_{\tilde{t}_L}^2
+\delta m_b^2-\delta m_t^2,\quad
\delta m_{\tilde{b}_L}^2|_{\rm OS4} =
\Sigma_{\tilde{b}_L}(m_{\tilde{b}_L}^2).
\end{equation}
The difference between the sfermion sector counterterms in the 2HDM
and the MSSM is thus contained in the mass shift
\begin{equation}
\Delta m_{\tilde{b}_L}^2 = \Sigma_{\tilde{b}_L}(m_{\tilde{b}_L}^2)-
\left(\delta m_{\tilde{t}_L}^2
+\delta m_b^2-\delta m_t^2\right).
\end{equation}
It turns out that this mass shift is only UV-finite if {\em all}
gauge-less limit relations are used, including $M_h=0$. Accordingly,
the MSSM result (\ref{MSSMRes})
is only finite if all relations in (\ref{glessrelations}) are
employed. The 2HDM result with on-shell renormalization of all four
sfermion masses is of course UV-finite for all choices of $M_h$.

\subsection{Incorporating the Higgs boson mass into the MSSM result}

The best way to take into account the non-vanishing value of $M_h$ as
much as possible is to consider the combination
\begin{equation}
\Delta\rho^{(\tilde{q},\tilde{H})}_{\rm MSSM}(M_h=0)+
\left(
\Delta\rho^{(\tilde{q},\tilde{H})}_{\rm 2HDM}(M_h)-
\Delta\rho^{(\tilde{q},\tilde{H})}_{\rm 2HDM}(M_h=0)
\right),
\label{FinalResult1}
\end{equation}
where the relations (\ref{Mh0nonzero}) are used everywhere. As
explained above, all three terms are individually UV-finite. Since the
difference between $\Delta\rho^{(\tilde{q},\tilde{H})}_{\rm MSSM}$ and
$\Delta\rho^{(\tilde{q},\tilde{H})}_{\rm 2HDM}$ is confined to the
$\tilde{t}/\tilde{b}$ counterterms, the combination
(\ref{FinalResult1}) can be written as
\begin{equation}
\Delta\rho^{(\tilde{q},\tilde{H})}_{\rm 2-Loop}(M_h)+
\Delta\rho^{(\tilde{q},\tilde{H})}_{\rm  t/b-cts}(M_h)+
\Delta\rho^{(\tilde{q},\tilde{H},\rm OS4)}_{\rm
  \tilde{t}/\tilde{b}-cts}(M_h)+
\left[
\Delta\rho^{(\tilde{q},\tilde{H},\rm OS3)}_{\rm
  \tilde{t}/\tilde{b}-cts}(M_h=0)
-
\Delta\rho^{(\tilde{q},\tilde{H},\rm OS4)}_{\rm
  \tilde{t}/\tilde{b}-cts}(M_h=0)
\right].
\label{FinalResult2}
\end{equation}
The first three terms correspond to the MSSM calculation where $M_h$
is set to its true, non-vanishing value but where all sfermion masses
are renormalized by on-shell conditions instead of using the mass
relation imposed by supersymmetry. 
The term in the square brackets is proportional to the mass shift
$\Delta m_{\tilde{b}_L}^2(M_h=0)$
that restores the necessary supersymmetry mass relation. It is only
here that $M_h=0$ has to be employed.

\section{NUMERICAL RESULTS}

\subsection{Results for different values of the supersymmetry
  parameters}

\begin{figure}[t]
\begin{picture}(600,120)
\epsfxsize=15cm
\put(20,0){\epsfbox{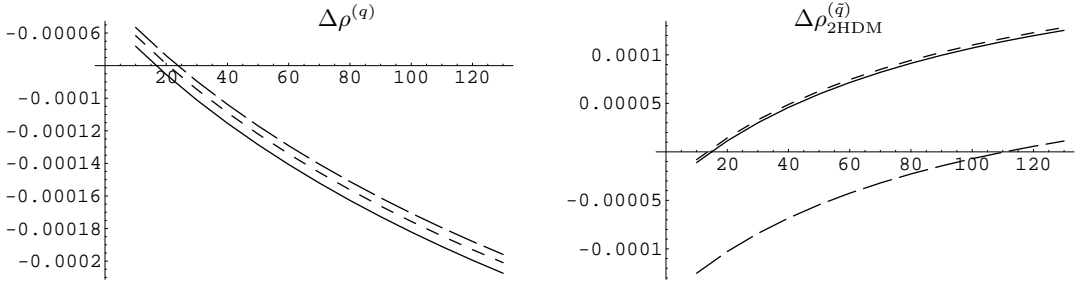}}
\put(150,110){$\Delta\rho^{(q)}$}
\put(330,110){$\Delta\rho^{(\tilde{q})}_{\rm 2HDM}$}
\end{picture}
\caption{$\Delta\rho^{(q)}$ and
$\Delta\rho^{(\tilde{q}_{\rm 2HDM})}$ as functions of $M_h$ for 
$M_{\rm SUSY}=400$ GeV, $A_t=1250$ GeV, $\tan\beta=2$,
  $\mu=500$ GeV and $M_A=$150 GeV (short-dashed), 300 GeV (full line),
  1000 GeV (long-dashed). This scenario involves a light stop with
  $m_{\tilde{t_1}}=120$ GeV.}
\label{fig:Mheffect}
\end{figure}

Fig.\ \ref{fig:Mheffect} demonstrates that it is in general important
to take into 
account the true value of the Higgs boson mass $M_h$. In the left and
right panels of Fig.\ \ref{fig:Mheffect}, $\Delta\rho^{(q)}$ and
$\Delta\rho^{(\tilde{q})}_{\rm 2HDM}$, which is the only $M_h$-dependent
term in eq.\ (\ref{FinalResult1}), are shown as functions of $M_h$ in
a scenario with a light stop (see caption). For both the fermion and
the sfermion loop contributions the difference of setting $M_h=0$ or
$M_h=$\order{100 \gev} amounts to more than $10^{-4}$.

\begin{figure}
\begin{picture}(600,150)
\epsfxsize=8cm
\put(10,0){\epsfbox{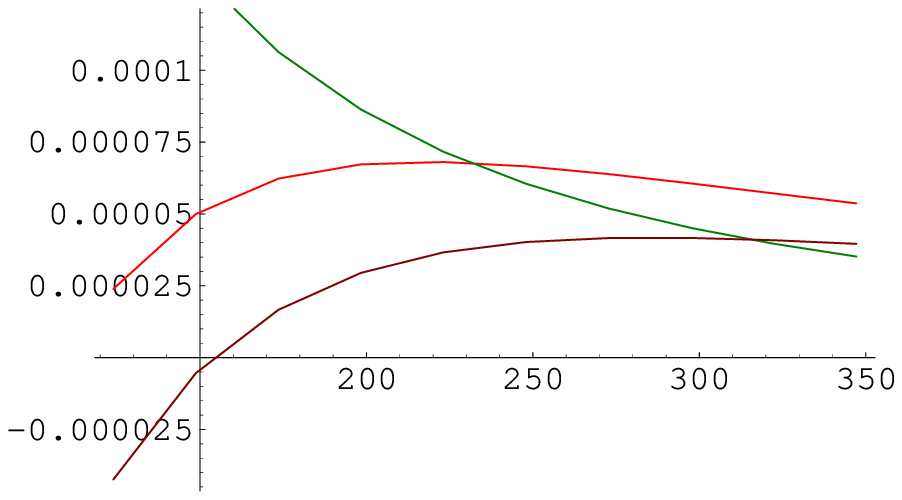}}
\put(90,51){(1)}
\put(90,117){(2)}
\put(90,82){(3)}
\epsfxsize=8cm
\put(275,0){\epsfbox{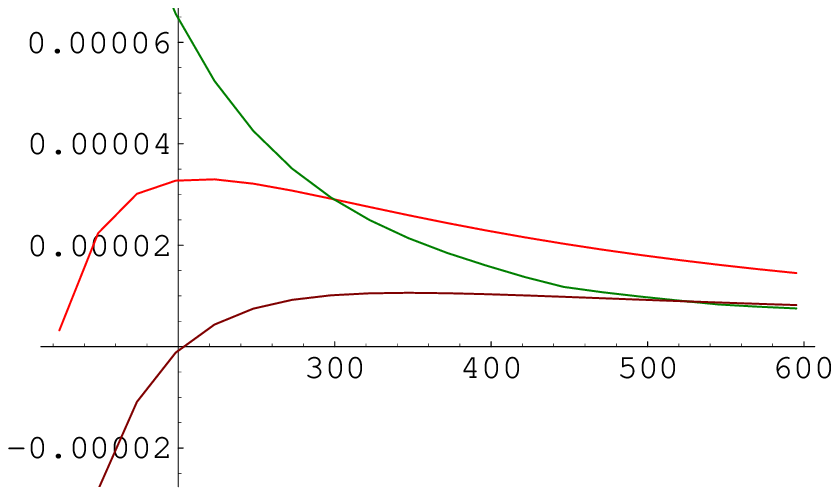}}
\put(330,53){(1)}
\put(330,124){(2)}
\put(330,93){(3)}
\end{picture}
\caption{Results for $\Delta\rho^{(\tilde{q})}(M_h=120\gev)$
  (1,brown), $\Delta\rho^{(\tilde{H})}$ (2,green), and, for comparison,
  $\Delta\rho^{(\tilde{q})}(M_h=0)$ (3,red), shown as functions of the
  common sfermion mass $M_{\rm  SUSY}$. In the left panel, the
  $A_{t,b}=0$; in the right panel, $A_{t,b}$ are chosen such that the
  ratios $M_{\rm SUSY}:A_t:A_b$ are as in the SPS1a scenario. The
  remaining supersymmetry   parameters are 
  always set   to the values of the SPS1a scenario.}  
\label{fig:SPSPlots}
\end{figure}

\begin{figure}
\begin{picture}(600,150)
\epsfxsize=8cm
\put(10,0){\epsfbox{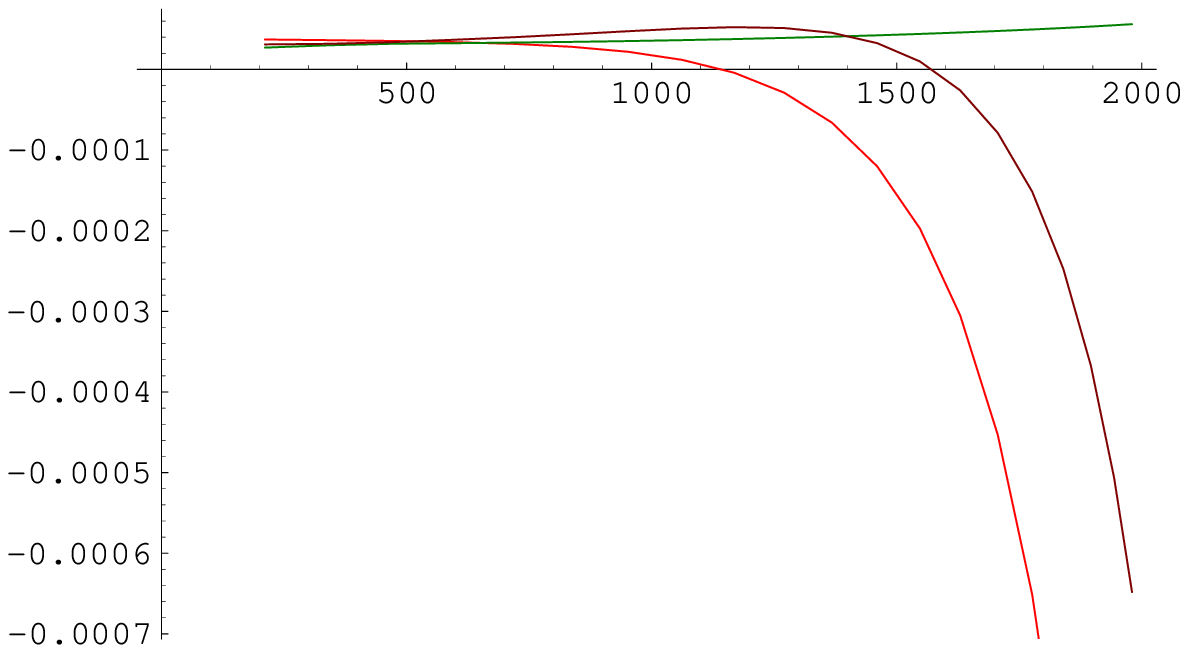}}
\put(215,80){(1)}
\put(215,130){(2)}
\put(190,30){(3)}
\epsfxsize=8cm
\put(275,0){\epsfbox{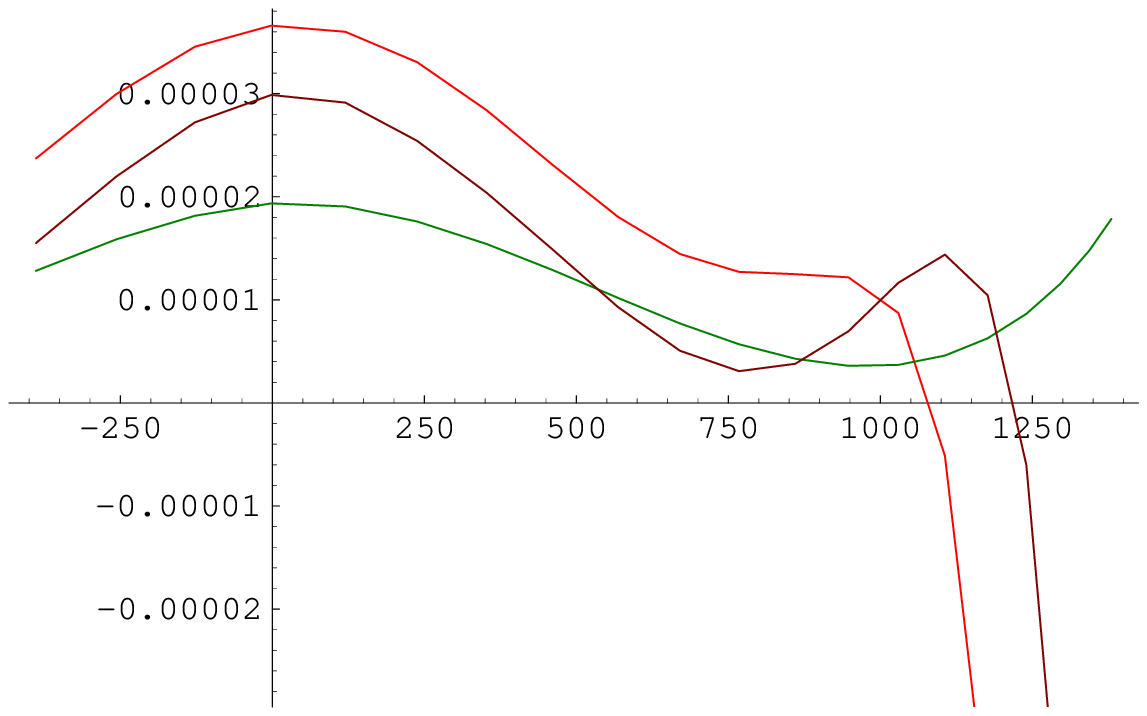}}
\put(455,97){(1)}
\put(492,102){(2)}
\put(405,102){(3)}
\end{picture}
\caption{Results for $\Delta\rho^{(\tilde{q})}(M_h=120\gev)$
  (1,brown), $\Delta\rho^{(\tilde{H})}$ (2,green), and, for comparison,
  $\Delta\rho^{(\tilde{q})}(M_h=0)$ (3,red). In the left panel, the
  results are shown as functions of the $\mu$-parameter for
  $\tan\beta=40$, such that for large $\mu$ one sbottom becomes light;
  in the right panel, the results are shown as 
  functions of $A_t$. The remaining supersymmetry parameters are
  always set to the values of the SPS1a scenario.}  
\label{fig:SPSPlots2}
\end{figure}

In the remainder we focus on the numerical results from the classes
$(\tilde{q},\tilde{H})$ (for class $(q)$ see Ref.\
\cite{drMSSMgf2A}). In Figs.\ \ref{fig:SPSPlots},~\ref{fig:SPSPlots2}
the results for 
$\Delta\rho$ as defined in eqs.\
(\ref{FinalResult1},\ref{FinalResult2}) are shown, split up into the
contributions for the individual classes. 
We choose several representative values for the supersymmetry
parameters as described in the captions, basically always starting
from the SPS1a scenario \cite{SPS} and 
varying one or two of the parameters. In the class $(\tilde{q})$
contributions the Higgs boson mass $M_h$ is set to either 120 GeV,
which is a good approximation for the true, loop-corrected value of
$M_h$, or to zero; the class $(\tilde{H})$ contributions are
$M_h$-independent.

According to eq.\ (\ref{precobs1}) a contribution of
 $\Delta\rho=10^{-4}$ leads to shifts
\begin{equation}
\delta M_W = 6{\mev},\quad
\delta\sweff = -3\times10^{-5}.
\end{equation}
We find that contributions of this order of magnitude are possible, in
particular for strong sfermion mixing (large values of $\mu$ or $A_t$
in Fig.\ \ref{fig:SPSPlots2}) but also for a light common sfermion
mass parameter $M_{\rm SUSY}$ as in Fig.\ \ref{fig:SPSPlots}.


\subsection{Different renormalization schemes and estimate of
  remaining theoretical uncertainty}

\begin{figure}
\begin{picture}(600,150)
\epsfxsize=8cm
\put(0,0){\epsfbox{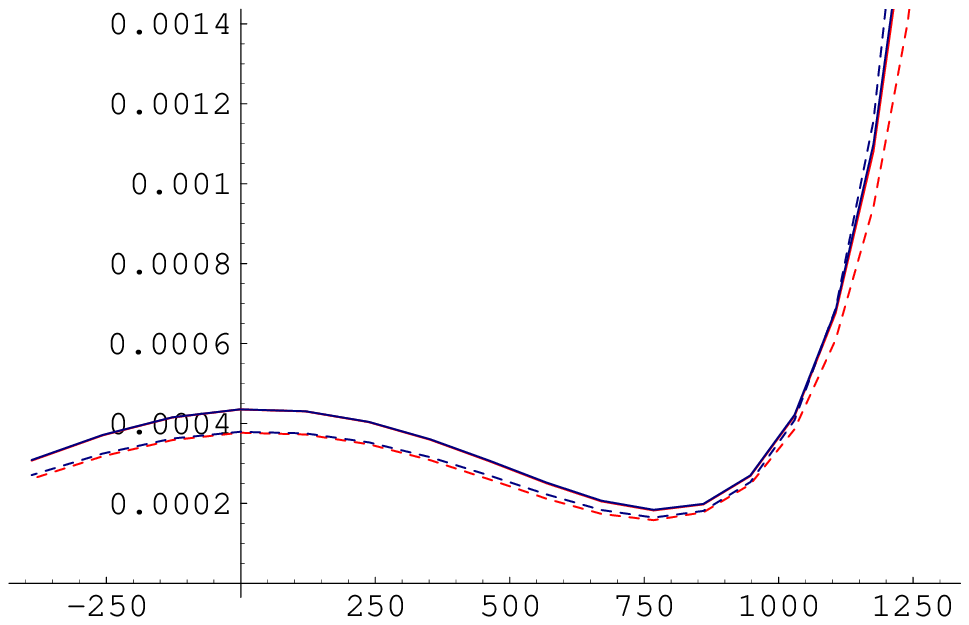}}
\epsfxsize=8cm
\put(275,0){\epsfbox{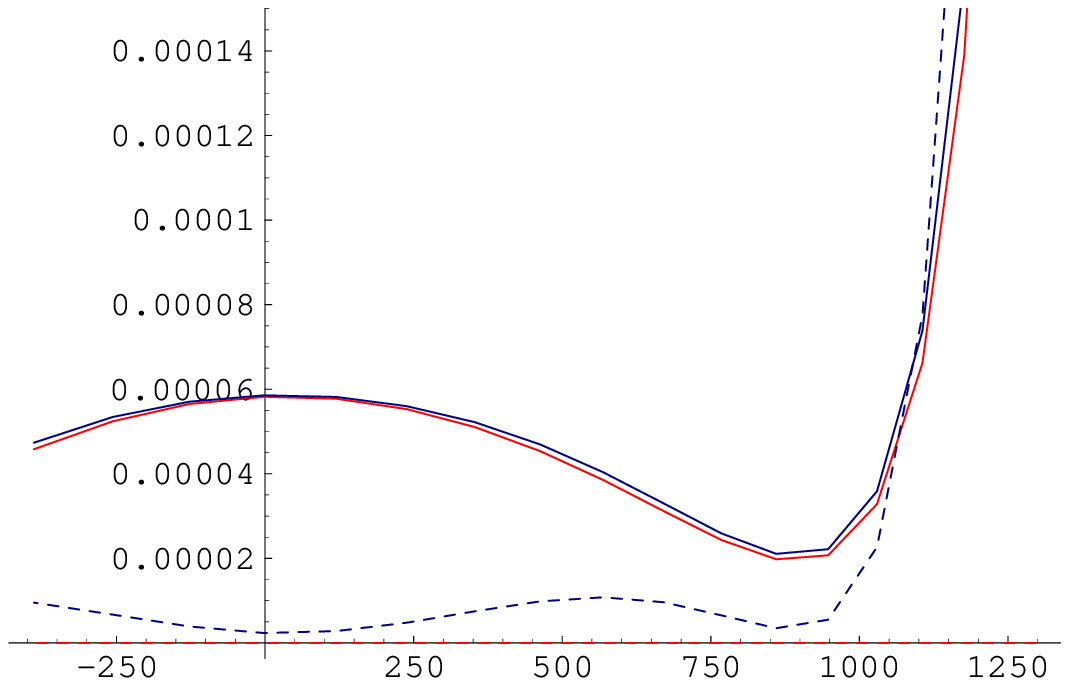}}
\put(390,37){(2)}
\put(380,23){(1)}
\put(400,49){(3)}
\end{picture}
\caption{The sfermion loop one-loop result $\Delta\rho^{(\rm 1L)}$
  (dashed) and the sum 
  $\Delta\rho^{(\rm 1L)}+\Delta\rho^{(\tilde{q},\tilde{H})}$ (full
  lines) in the 
  on-shell (blue) and the $\overline{\rm DR}$ (red) renormalization
  scheme. 
  The parameters are as in the right panel of Fig.\
  \ref{fig:SPSPlots2}. The left panel shows the full results, where
  the differences between the curves are hardly visible; the
  right panel shows the differences of the results to
  $\Delta\rho^{(\rm 1L)}|_{\overline{\rm DR}}$, i.e.\ (1) shows
  $\Delta\rho^{(\rm 1L)}|_{\rm OS}-\Delta\rho^{(\rm
  1L)}_{\overline{\rm DR}}$, (2,3) show  
$(\Delta\rho^{(\rm 1L)}+
\Delta\rho^{(\tilde{q},\tilde{H})})_{\overline{\rm DR},{\rm OS}}-
\Delta\rho^{(\rm  1L)}_{\overline{\rm DR}}$. For simplicity, $M_h=0$
  is used here, but the conclusions do not change for non-vanishing
  $M_h$.}
\label{fig:RenScheme}
\end{figure}

So far we have chosen the on-shell renormalization scheme for the
independent sfermion masses $m_{\tilde{t}_{1,2}}$ and
$m_{\tilde{b}_2}$. In Fig.\ \ref{fig:RenScheme} the on-shell results
are compared with the results in the $\overline{\rm DR}$ scheme,
where the counterterms for the soft supersymmetry breaking parameters
in the sfermion mass matrices are defined as pure divergences. This
$\overline{\rm DR}$ scheme implies non-vanishing finite parts of the
sfermion mass counterterms, which read, in the
case of vanishing left-right mixing, $\delta m_{\tilde{f}_i}^2|_{\rm
    fin-part}=\delta m_f^2|_{\rm fin-part}$.
The comparison of the two renormalization schemes is interesting in
order to assess the numerical stability of the result and the
intrinsic theoretical uncertainty of the two-loop contributions. 

We find that the difference between the on-shell and $\overline{\rm
  DR}$ results for the sfermion loop contributions at the one-loop
level is of the order $10^{-5}$ to 
$10^{-4}$ for large sfermion mixing. At the two-loop level this
renormalization-scheme dependence is significantly reduced to well
below $10^{-5}$. We have checked that this result is general and not
restricted to the particular parameter choice in Fig.\
  \ref{fig:RenScheme}.

\newpage

In conclusion, we have evaluated the 
\order{\alpha_t^2}, \order{\alpha_t  \alpha_b}, \order{\alpha_b^2}
corrections to $\Delta\rho$ and thus to the EWPO $M_W$ and $\sweff$ in
the MSSM. Although the gauge-less limit is necessary and leads to the
tree-level relation $M_h=0$, we have shown that the true MSSM Higgs
boson mass can be taken into account. The numerical values of the
class $(\tilde{q},\tilde{H})$ contributions (\ref{FinalResult1}), 
(\ref{FinalResult2}) to $\Delta\rho$ can amount
to $10^{-4}$, corresponding to shifts
$\delta M_W = 6{\mev},
\delta\sweff = -3\times10^{-5}$.
The comparison of the two renormalization schemes shows that the
inclusion of the two-loop result  in the MSSM-prediction for
$\Delta\rho$ and the 
EWPO leads to a significantly improved accuracy. The residual
theoretical uncertainty due to unknown three-loop corrections of
\order{\alpha_{t,b}^3} is well below the foreseen
experimental resolution achievable at the GigaZ option of a linear
$e^+e^-$ collider.

\end{document}